\documentclass[doublespacing]{elsart}

\usepackage{amssymb}
\usepackage{epigraph}
\usepackage{epsfig}

\newtheorem{Cor}{Corollary}
\newtheorem{Prop}{Proposition}
\newtheorem{Ex}{Example}

\newcommand{\be}{\begin{equation}}
\newcommand{\ee}{\end{equation}}
\newcommand{\lb}{\label}
\newcommand{\ol}{\overline}
\newcommand{\ba}{{\bf a}}
\newcommand{\bb}{{\bf b}}
\newcommand{\bF}{{\bf f}}
\newcommand{\bi}{{\mbox{\boldmath ${\i}$}}}
\newcommand{\bj}{{\mbox{\boldmath ${\j}$}}}
\newcommand{\bk}{{\bf k}}
\newcommand{\Be}{{\bf e}}
\newcommand{\br}{{\bf r}}
\newcommand{\bu}{{\bf u}}
\newcommand{\bx}{{\bf x}}
\newcommand{\bz}{{\bf z}}
\newcommand{\bA}{{\bf A}}
\newcommand{\bB}{{\bf B}}
\newcommand{\bE}{{\bf E}}
\newcommand{\bR}{{\bf R}}
\newcommand{\sign}{{\rm sign}\,}

\newcommand{\cD}{{\mathcal D}}
\newcommand{\bomega}{{\mbox{\boldmath $\omega$}}}
\newcommand{\bepsilon}{{\mbox{\boldmath $\varepsilon$}}}
\newcommand{\bxi}{{\mbox{\boldmath $\xi$}}}
\newcommand{\grad}{{\mbox{\boldmath $\nabla$}}}
\newcommand{\bdot}{{\mbox{\boldmath $\cdot$}}}
\newcommand{\btimes}{{\mbox{\boldmath $\times$}}}
\newcommand{\bzed}{{\mbox{\boldmath $0$}}}

\begin{document}

\begin{frontmatter}

\title{The Breakdown of Alfv\'{e}n's Theorem in Ideal Plasma Flows}
\author{Gregory L. Eyink and Hussein Aluie}
\ead{eyink@ams.jhu.edu}
% \ead[url]{http://ams.jhu.edu/~eyink}
\address{
Department of Applied Mathematics \& Statistics\\
The Johns Hopkins University \\
3400 North Charles Street \\
Baltimore, MD 21218-2682\\
\& \\
Center for Nonlinear Science\\
Los Alamos National Laboratory\\
Los Alamos, NM 87545
}

\begin{abstract}
This paper presents both rigorous results and physical theory on the breakdown
of magnetic flux conservation for ideal plasmas, by nonlinear effects. Our
analysis
is based upon an effective equation for magnetohydrodynamic (MHD) modes at
length-scales $>\ell,$ with smaller scales eliminated, as in
renormalization-group
methodology. We prove that flux-conservation can be violated for an arbitrarily
small
length-scale $\ell,$ and in the absence of any non-ideality, but only if
singular current
sheets and vortex sheets both exist and intersect in sets of large enough
dimension.
This result gives analytical support to and rigorous constraints on theories
of fast turbulent reconnection. Mathematically, our theorem is analogous to
Onsager's
result on energy dissipation anomaly in hydrodynamic turbulence. As a physical
phenomenon, the breakdown of magnetic-flux conservation in ideal MHD is similar
to the decay of magnetic flux through a narrow superconducting ring, by
phase-slip
of quantized flux lines. The effect should be observable both in numerical MHD
simulations and in laboratory  plasma experiments at moderately high magnetic
Reynolds numbers.
\end{abstract}


\end{frontmatter}

\section{Introduction}

\setlength{\beforeepigraphskip}{.5\baselineskip}
\epigraphwidth=250pt
\epigraph{{\it In view of the infinite conductivity, every motion
  (perpendicular to the field) of the liquid in relation to the
  lines of force is forbidden because it would give infinite
  eddy currents. Thus the matter of the liquid is ``fastened''
  to the lines of force...}}{\textsc{H. Alfv\'{e}n (1942)}}

It is a fundamental result for an ideal plasma, or perfectly conducting
fluid, that magnetic lines of force are ``frozen-in'' and move with the
fluid. This fact was first pointed out by Hannes Alfv\'{e}n in 1942
\cite{Alfven42}, in the quote above. About the same time a somewhat stronger
result was also observed, that the magnetic flux through a surface moving
with a perfectly conducting fluid is conserved. For a good historical
review of this early work, see \cite{Stern66}. These properties of magnetic
field lines exactly parallel the corresponding properties established by
Helmholtz \cite{Helmholtz:1858} for vortex lines in an ideal, inviscid fluid.
The results for ideal magnetohydrodynamics (MHD)---the ``frozen-in''
property of field lines and the conservation of magnetic flux---are often
referred to together as ``Alfv\'{e}n's Theorem''.

One important consequence of these theorems is that magnetic field
lines in a perfectly conducting fluid cannot change their topology.
In particular, the reconnection of crossed magnetic field lines is
forbidden (e.g. see \cite{Axford84}). This poses a bit of a paradox,
however. If magnetic field lines are not able to pass through each
other more or less freely, then one would expect them to form a
complicated tangle that would strongly impede plasma motion, or even
thwart it altogether \cite{Lazarian05}. The theory of ideal plasmas
would then be closely analogous to the statistical mechanics of
rubber elasticity, where the entanglements of polymer chains
imply an infinite set of topological constraints, of which the Gauss linking
number is just the simplest \cite{DeamEdwards76,GoldbartGoldenfeld04}.
However, despite the fact that near-ideal conditions
hold in a wide variety of astrophysical situations
(interstellar space, the solar corona, etc.), the behavior of
these plasmas is not at all ``rubber-like'' but instead essentially
fluid-dynamical. The implication is that reconnection of magnetic field
lines occurs at rates that are nearly independent, or even independent,
of the value of the resistivity.  Explaining this phenomenon of
``fast reconnection'' is a well-known problem of plasma physics and
astrophysics \cite{Biskamp00,PriestForbes00}, with important implications
for understanding dynamo action \cite{Parker92}, solar flares and
coronal mass ejections \cite{Dere96}, etc. Since Alfv\'{e}n's Theorem
prohibits reconnection, any fundamental theory of fast reconnection
must explain also the breakdown of those ideal MHD results.

Small Ohmic resistivity (or other non-ideality) implies high magnetic
Reynolds numbers, so that laminar solutions of near-ideal MHD
equations will be unstable and the plasma motion will generally
be turbulent. Theories of fast turbulent reconnection have been
proposed \cite{LazarianVishniac99,LazarianVishniac00}, which predict
magnetic-flux reconnection rates that are completely independent
of the resistivity. This implies implicitly the violation of  Alfv\'{e}n's
Theorem
under ideal conditions in turbulent plasmas. Such a breakdown of classical
conservation laws of the fluid equations under turbulent conditions is not
unprecedented. For example, it is well-known that energy is not conserved in
the
limit of small viscosity for hydrodynamic turbulence, as observed both in
laboratory
experiments \cite{Dryden43,Sreenivasan84,Cadotetal97,Pearsonetal02}
and high-resolution numerical simulations
\cite{Sreenivasan98,Kanedaetal03}. A fundamental explanation for
this phenomenon was proposed in 1949 by Lars Onsager \cite{Onsager49},
who showed that solutions of the ideal incompressible Euler equations
do not need to conserve energy if they are sufficiently singular.
More precisely, Onsager showed that, if the turbulent velocity field
is not differentiable in space but only H\"{o}lder continuous with an
exponent $\leq 1/3,$ then the observed rate of energy dissipation could be
explained without any viscosity.
See also \cite{Eyink94,Constantinetal94,DuchonRobert00}, and
\cite{EyinkSreenivasan06} for a recent review.
Onsager's prediction of near-singularities in turbulent velocity fields
with H\"{o}lder index $\leq 1/3$ has been confirmed by analysis of
high-Reynolds number data from experiments and numerical simulations
\cite{Muzyetal91,Arneodoetal95,KestenerArneodo04}.

Recently, one of us (G.E.) has extended Onsager's results
on inviscid energy disssipation to the breakdown of the
Helmholtz-Kelvin Theorem in hydrodynamic turbulence at
high Reynolds number \cite{Eyink06a,Eyink06b}. The result
proved there was that conservation of circulations can
break down if the velocity field has near-singularities
with H\"{o}lder exponent $\leq 1/2,$ and it was conjectured
on this basis that there would be a ``cascade of circulations''
for fluid turbulence in three spatial dimensions. This prediction
has been confirmed by a high-resolution numerical simulation
\cite{Chenetal06}. The main purpose of the present paper is
to establish corresponding mathematical results on the breakdown
of Alfv\'{e}n's Theorem for singular solutions of the
ideal MHD equations. Despite the similarity of Alfv\'{e}n's result
to the Helmholtz-Kelvin Theorem in hydrodynamics, there are
important differences between the results obtained here
and those established in \cite{Eyink06a,Eyink06b}. It turns
out that it is much harder to violate conservation of
magnetic flux than it is to violate conservation of circulation.
H\"{o}lder continuity exponents that are only moderately small,
and far from the most singular behavior in hydrodynamic turbulence,
can produce breakdown of the Helmholtz-Kelvin Theorem. The main
result proved here can be stated succinctly (but somewhat
imprecisely) as follows: {\it Alfv\'{e}n's Theorem can break
down in ideal (or near-ideal) magnetohydrodynamics only due to
intersecting current sheets and vortex sheets.} The latter are
the most singular structures expected to occur in plasma turbulence
and only these, acting together, can lead to violation of magnetic
flux conservation at a rate which is independent of resistivity.
This is a quite striking difference from the hydrodynamic case.

The contents of this paper are as follows: In the following
Section 2 we briefly review the formal statement of Alfv\'{e}n's
Theorem, its derivations, and its implications for magnetic line
reconnection. In the next Section 3 we present the ``filtering
approach'',  which is the basis of our entire analysis and explain
its relation to theory of distributions (or generalized functions),
renormalization group theory, and large-eddy simulation.
In Section 4 we prove the main results of the paper. In
Section 5 we consider the physical possibilities for breakdown
of Alfv\'{e}n's Theorem and for ideal reconnection, based
upon our rigorous results. In Section 6 we discuss the possibilities
for a ``cascade of magnetic flux'' in MHD turbulence and fast
turbulent reconnection. Finally, in Section 7 we restate succinctly
our main conclusions.

\section{Alfv\'{e}n's Theorem on Magnetic Flux Conservation}

We review here briefly some standard derivations of Alfv\'{e}n's Theorem(s).
These are based only upon the homogeneous Maxwell equations
\be \grad\bdot\bB=0,\,\,\,\, \partial\bB/\partial t +\grad\btimes\bE=0
\lb{homo-Maxwell} \ee
and the general Ohm's law
\be  \bE + \bu\btimes\bB =\bR, \lb{general-Ohm} \ee
where $\bR$ represents an arbitrary non-ideality and $\bu$ is any
time-dependent
velocity field, not necessarily the solution of any fluid equation. For
simplicity,
we shall assume here that the velocity is incompressible, $\grad\bdot\bu=0,$
although this is not crucial for the discussion.

The magnetic flux as a Lagrangian variable is defined by
\be \Phi(S,t) = \int_{S(t)} \,\bB(t)\bdot d\bA \lb{flux} \ee
where the initial surface $S$ is smooth and oriented, and $S(t)$ is the surface
at later times advected by the velocity field $\bu.$ The standard proof of
flux conservation uses the easily verified result that
\be (d/dt) \Phi(S,t) =  \int_{S(t)} \,\left[ \partial \bB/\partial t
-\grad\btimes
     (\bu\btimes\bB) \right]\bdot d\bA, \lb{standard} \ee
where the second term in the square bracket represents the change in the flux
due to motion of the surface. For example, see \cite{Stern66} or
\cite{Chandrasekhar61},
section \S 38. Taking the curl of Ohm's law (\ref{general-Ohm}) and
using Faraday's law (\ref{homo-Maxwell}), gives
\be  \partial\bB/\partial t =\grad\btimes
     (\bu\btimes\bB) -\grad\btimes\bR. \lb{general-Faraday} \ee
Substituting this result into (\ref{standard}) and using Stokes Theorem gives
\be (d/dt)\Phi(S,t) = -\oint_{C(t)}\, \bR\bdot d\bx \lb{Alfven-flux} \ee
where $C(t)=\partial S(t)$ is the boundary curve of the advected surface
$S(t).$ For  $\grad\btimes\bR=\bzed,$ and in particular for vanishing
non-ideality, $\bR=\bzed,$ flux conservation immediately follows.

It is a simple consequence of flux-conservation that magnetic
flux-tubes---whose
surface normal is everywhere perpendicular to the magnetic field---are material
surfaces.
Since magnetic field lines are intersections of magnetic flux tubes, they must
likewise be
material lines. Thus, the ``frozen-in'' property of magnetic field lines is a
direct consequence
of flux-conservation.  More formally, the mathematical condition for line
preservation is
\cite{Stern66}:
$$ \left[ \partial \bB/\partial t -\grad\btimes
     (\bu\btimes\bB) \right]\btimes\bB=\bzed. $$
This is the exact analogue of the Helmholtz-Zorawski condition for preservation
of
vortex lines under fluid advection \cite{Zorawski00}. Thus, the condition on
the non-ideality
for  ``frozen-in'' magnetic field lines is $(\grad\btimes\bR)\btimes
\bB=\bzed,$ which
is weaker than the condition $\grad\btimes\bR=\bzed$ for flux-conservation.

There are other derivations of the Alfv\'{e}n Theorem that reveal more of its
geometric and dynamic significance. The equation (\ref{general-Faraday}) with
$\grad\btimes\bR=\bzed$ is equivalent to the equation
$$  \partial_t F + {\mathcal L}_\bu F=0 $$
where $F=F_{ij}\,dx_i\wedge dx_j$ is the spatial magnetic 2-form and
${\mathcal L}_\bu$ is the Lie-derivative for the vector field $\bu.$
Flux-conservation
then follows immediately from the Lie-derivative Theorem \cite{Abrahametal83}.
Another derivation can be based upon the Hamiltonian formulation of ideal
magnetohydrodynamics, which possesses an infinite-dimensional gauge
symmetry group corresponding to relabelling of fluid particles. In this
framework,
conservation of magnetic flux for all smooth, oriented surfaces $S$ is a
consequence
of Noether's theorem for the relabelling symmetry \cite{Ruban99}.

If the fluid flow is continuous, then the ``frozen-in'' property forbids any
change of
magnetic line topology, such as reconnection. A more formal connection with
Alfv\'{e}n's Theorem appears in certain theories of $\bB\neq\bzed $ magnetic
reconnection
\cite{Schindleretal88,HesseSchindler88,Schindler95}, based upon the
{\it magnetic loop-voltage}
\be V_L = \oint_L \bR\bdot d\bx, \lb{loop-voltage} \ee
where $L$ is a magnetic field line (which may be closed at infinity). This
integral
is the same type of voltage which appears in the flux balance
(\ref{Alfven-flux}),
breaking flux-conservation, and it vanishes under the same condition
$\grad\btimes\bR
=\bzed$ for which Alfv\'{e}n's Theorem is valid.  The magnetic lines $L$
passing
through the region of non-ideality that have extremal values of the
loop-voltage are called
{\it reconnection} (or {\it separator}) {\it lines}.  Under certain assumptions
it can
proved that these lines do not undergo reconnection themselves but drive the
reconnection
of neighboring lines \cite{Schindleretal88,HesseSchindler88,Schindler95}. Thus,
the
loop-voltages that produce $\bB\neq \bzed$ reconnection have the same origin as
the voltages
from any non-ideality that violates Alfv\'{e}n's Theorem.

\section{The Filtering Approach and Large-Scale Flux Balance}

The proofs of Alfv\'{e}n's Theorem discussed in the preceding
section all assume, implicitly, that the solutions $\bu$ and
$\bb$\footnote{Hereafter we use Alfv\'{e}n velocity variables $\bb=
\bB/\sqrt{4\pi\rho_0},$ with $\rho_0$ the plasma density, rather than
magnetic field variables $\bB $ and the corresponding variable
$\Be=\bE/\sqrt{4\pi\rho_0}$ rather than the electric field $\bE$.}
of the MHD equations remain smooth in the limit where
the resistivity $\eta$ (or other non-ideality) tends to zero.
However, those proofs can break down if the solutions become
singular in that limit. To see how this can occur, let us
consider the ideal Ohm's law
\be  \Be + \bu\btimes\bb =\bzed, \lb{ideal-Ohm} \ee
and its curl, using Faraday's law,
\be  \partial \bb/\partial t =\grad\btimes
     (\bu\btimes\bb), \lb{ideal-Faraday} \ee
in the case that $\bu$ and $\bb$ are singular. In that case,
equation (\ref{ideal-Faraday}), in particular, is not meaningful
in a naive sense, because the classical derivative of a non-smooth
function is ill-defined.  The natural way to interpret equation
(\ref{ideal-Faraday}) is in the sense of distributions, which means
that it must be multiplied by smooth test functions $\varphi(\bx)$
and integrated over $\bx,$ allowing the curl to be shifted to
the test function. (We assume here, for simplicity, that the
time-dependence of the solutions is smooth so that test
functions $\varphi(\bx,t)$ in both variables are not required.)
Because equation (\ref{ideal-Faraday}) is quadratically nonlinear,
it is not hard to see that it is well-defined, in the sense
of distributions, whenever $\bu,\bb\in L^2,$ i.e. are
square-integrable, or, physically, have finite energy.
In principle, all test functions in a suitable space---e.g.
the set $C^\infty_c(\mathbb{R}^3)$ of infinitely-differentiable
functions with compact support---should be considered.
Fortunately, it is not necessary to consider {\it all} of the
test functions in the space, but only a subset defined by
\be \varphi_{\ell,\bx}(\bx')=\ell^{-3}G(\frac{\bx'-\bx}{\ell})
      \lb{test-filter} \ee
for all $0<\ell<\ell_0$ and $\bx\in \mathbb{R}^3,$ where $G$ is
a particular function that satisfies
\be G\in C^\infty_c(\mathbb{R}^3),\,\,\,\, G\geq 0,\,\,\,\,
   \int d\br \, G(\br)=1 \lb{kernel} \ee
 One may also substitute here a
condition of rapid decay of $G(\br)$ at large $|\br|,$ faster than
any power. For these standard facts of distribution theory, see
\cite{Friedlander98,Hormander85}, for example.

There is a more physical  way to explain this formulation of the
equations (\ref{ideal-Ohm}) and (\ref{ideal-Faraday}), which makes
clear how Alfv\'{e}n's Theorem may be broken. Integrating the solutions
$\bu,\bb$ with respect to the test function (\ref{test-filter}) yields
\be \ol{\bu}_\ell(\bx) = \int d\br \, G_\ell(\br) \bu(\bx+\br) \lb{filter-u}
\ee
and similarly for $\ol{\bb}_\ell(\bx),$ where
$G_\ell(\br)=\ell^{-3}G(\br/\ell).$
The fields $\ol{\bu}_\ell,\ol{\bb}_\ell$ are ``coarse-grained'' at length-scale
$\ell $ or low-pass filtered, retaining information only from scales $>\ell.$
These filtered fields satisfy a modified form of Ohm's law
\be \ol{\Be}_\ell + \ol{\bu}_\ell\btimes\ol{\bb}_\ell = -\bepsilon_\ell
                        \lb{filter-Ohm} \ee
where the {\it subscale electromotive force (EMF)}
\be   \bepsilon_\ell = \ol{(\bu\btimes\bb)}_\ell
-\ol{\bu}_\ell\btimes\ol{\bb}_\ell
         \lb{filter-EMF} \ee
provides an effective non-ideality. The subscale EMF is also sometimes
referred to as {\it turbulent EMF} and, in fact, the present scheme is closely
related to the so-called  ``filtering approach'' used in Large-Eddy Simulation
(LES) modelling of turbulent flows \cite{Germano92,MeneveauKatz00}.
In this approach the filtered version of equation (\ref{ideal-Faraday}),
\be  \partial \ol{\bb}_\ell/\partial t =\grad\btimes
     (\ol{\bu}_\ell\btimes\ol{\bb}_\ell)+\grad\btimes\bepsilon_\ell,
\lb{filter-Faraday} \ee
would be solved, together with similar filtered equations for the velocity,
by employing a closure model for the EMF term $\bepsilon_\ell.$

This same term breaks the validity of flux-conservation for the coarse-grained
fields. In fact, the standard derivations of Alfv\'{e}n's Theorem now imply
that
\be (d/dt)\ol{\Phi}_\ell(S,t)\equiv (d/dt)\int_{\ol{S}_\ell(t)}
\ol{\bb}_\ell(t)\bdot d\bA
       = \oint_{\ol{C}_\ell(t)} \bepsilon_\ell\bdot d\bx.  \lb{filter-Alfven}
\ee
Here our notations are the same as those in the preceding section,
except that  the surface $\ol{S}_\ell(t)$ and its boundary curve
$\ol{C}_\ell(t)$
are defined to be those advected by the filtered velocity $\ol{\bu}_\ell.$
This
``large-scale flux balance'' now contains a source/sink term which can violate
Alfv\'{e}n's Theorem. The physical origin of this phenomenon is an effective
``drift velocity''  $\Delta \ol{\bu}_\ell$ of the field-lines of the
large-scale  magnetic
field $\ol{\bb}_\ell$ relative to the plasma velocity $\ol{\bu}_\ell,$ induced
by
the subscale EMF.   This drift velocity is not uniquely defined, and its
component
$\Delta\ol{\bu}_\ell^{\|}$ parallel to the field lines is largely arbitrary,
but it
always contains a transverse component \cite{Newcomb58}:
\be \Delta\ol{\bu}_\ell^{\perp}
   =\bepsilon_\ell \btimes\ol{\bb}_\ell/|\ol{\bb}_\ell|^2.  \lb{drift-velocity}
\ee
A rather natural prescription to define a drift velocity $\Delta\ol{\bu}_\ell$
is
proposed in \cite{Boozer90}. Because of the additional normal velocity
component
(\ref{drift-velocity}),  the field lines now slip through the plasma and
Alfv\'{e}n's
``frozen-in'' property is violated at length-scale $\ell.$

The violation of Alfv\'{e}n's Theorem so far considered is, in some sense, not
``real''.
For example, the ideal MHD equations would gain an additional EMF even for a
smooth, laminar solution, if that were filtered at any length-scale $\ell>0.$
However,
for such solutions the subscale EMF would vanish rapidly in the limit
$\ell\rightarrow 0$
(see following section). {\it We define a real violation of Alfv\'{e}n's
Theorem
for an ideal MHD solution as one which persists in the limit $\ell\rightarrow
0.$}
This definition can be justified physically by a Renormalization Group (RG)
argument \cite{Goldenfeld92}.
%  for a lucid introduction to this subject.)
The dynamical equation (\ref{filter-Faraday}) for $\ol{\bb}_\ell$ is a
``renormalized''
equation, obtained by integrating out the high-wavenumber modes. In contrast
to the ``bare'' equation (\ref{ideal-Faraday}), it contains only observable
quantities.
An experiment can measure the velocity and magnetic fields, in fact,  only
down to a certain spatial resolution corresponding to a length-scale $\ell.$
If an experimentalist  were to attempt to verify magnetic flux conservation,
then he would be testing its validity for some coarse-grained fields
$\ol{\bu}_\ell,\ol{\bb}_\ell$ and not for the bare fields $\bu,\bb.$  Because
of the
subscale EMF, flux conservation would be violated to some extent for any
$\ell>0,$
but the experimentalist would say that flux conservation was verified if it
held with
increasing accuracy for finer resolutions $\ell.$ On the contrary, the
experimentalist
would be forced to say that flux conservation was violated if the effects of
the
turbulent EMF $\bepsilon_\ell $ did not vanish for arbitrarily small $\ell.$

Note that this violation, if present, is entirely an effect of the nonlinearity
and
not due to any  standard non-ideality, such as Ohmic resistivity or other
anomalous
transport coefficients, such as ambipolar diffusion
\cite{MestelSpitzer56,TassisMouschovias04}, Braginskii viscosity
\cite{Braginskii65,Schekochihinetal05},
etc.  The direct effect of such non-ideal terms will be negligibly small for a
sufficiently
large length-scale $\ell,$ much greater than the dissipation length set by
resistivity
or than internal plasma lengths, such as the ion gyroradius and ion skin depth.
For example, in the presence of Ohmic resistivity $\eta$, the large-scale
effective
equation (\ref{filter-Faraday}) would contain an additional Laplacian term
$$ \eta\bigtriangleup\ol{\bb}_\ell(\bx) = \eta \ell^{-2} \int d\br \,
                    (\bigtriangleup G)_\ell(\br) \bb(\bx+\br). $$
For any solution with finite magnetic energy, $\|\bb\|_2<\infty,$ the above
equation can be used to provide a bound $\|\eta\bigtriangleup\ol{\bb}_\ell\|_2
\leq ({\rm const.}) (\eta/\ell^2)\|\bb\|_2.$ Therefore, this term vanishes in
the limit
$\ell\rightarrow\infty$ with $\eta$ fixed, or  $\eta\rightarrow 0$ with $\ell$
fixed.
For a small ratio $\eta/\ell^2$ the Ohmic dissipation term is negligible in the
large-scale equation. Of course, for a sufficiently small length-scale
$\ell_d,$
these dissipative, non-ideal effects will be significant. At such small scales,
Alfv\'{e}n's Theorem will be violated due to the dissipative, non-ideal
effects.
However, at lengths $\ell\gg\ell_d,$ Alfv\'{e}n's Theorem can be violated by
the subscale EMF due to the nonlinearity. If both effects exist, then there is
no
range of  length-scales whatsoever where flux-conservation holds.

In the next section we investigate the properties of the solutions $\bu,\bb$
that permit a persistent effect of the nonlinearity for $\ell>\ell_d.$ We find
that rather strong singularities are required.

\section{Theorems on Flux Conservation}

We now prove several simple theorems on sufficient conditions for
conservation of magnetic flux, or, equivalently, necessary conditions
for the breakdown of flux conservation. A key tool for our analysis
is the following formula for the subscale EMF
\be \bepsilon_\ell(\bx) =
   \int d\br \,G_\ell(\br) \, \delta\bu(\br;\bx)\btimes\delta\bb(\br;\bx)
   - \int d\br \,G_\ell(\br)\, \delta\bu(\br;\bx) \,
   \btimes \int d\br \,G_\ell(\br) \, \delta\bb(\br;\bx)
\lb{EMF-delta} \ee
where $\delta\bu(\br;\bx)=\bu(\bx+\br)-\bu(\bx)$ is the velocity increment
at point $\bx$ for separation vector $\br$ and similarly for
$\delta\bb(\br;\bx).$
The formula (\ref{EMF-delta}) is easily verified by multiplying out
the factors and integrating over the separations.

The first point to recognize is that the subscale EMF vanishes nearly
everywhere in the limit as $\ell\rightarrow 0,$ under very minimal
conditions. For example, let us assume just that the total energy
(kinetic and magnetic) is finite:
\be E = \frac{1}{2}\left[\|\bu\|_2^2+\|\bb\|_2^2\right]<\infty, \lb{energy} \ee
Here $\|\bu\|_2^2=\int_\Lambda d\bx\,|\bu(\bx)|^2$ defines the
standard $L^2$ norm in the flow domain $\Lambda,$ and similarly
for $\|\bb\|_2^2.$ Then the following result holds:
\begin{Prop}
Let $\bu,\bb\in L^2.$ Then $\bepsilon_\ell
\longrightarrow^{\!\!\!\!\!\!\!\!\! L^1} 0 $
as $\ell\rightarrow 0.$
\end{Prop}
If the flow domain $\Lambda$ is infinite, it is more natural
to substitute the conditions that $\bu,\bb$ have locally finite
energy densities and then the convergence of $\bepsilon_\ell$
to zero is in the local $L^1$ sense.

{\it Proof of Proposition 1:} A standard density argument from real
analysis shows that the $L^2$-norms of the increments, $\|\delta\bu(\br)\|_2$
and $\|\delta\bb(\br)\|_2,$ are continuous in $\br$ and, in particular,
vanish as $\br\rightarrow\bzed.$ In fact, by the reverse triangle
inequality,
$$ |\|\delta\bu(\br)\|_2-\|\delta\bu(\br')\|_2|\leq
    \|\delta\bu(\br)-\delta\bu(\br')\|_2
    =\|\bu(\cdot+\br)-\bu(\bdot+\br')\|_2. $$
Since smooth functions are dense in $L^2,$ there exists a smooth
function $\bu^*$ so that $\|\bu-\bu^*\|_2<\epsilon$ for any
$\epsilon>0.$ Thus, by triangle inequality,
$$ \|\bu(\cdot+\br)-\bu(\bdot+\br')\|_2 \leq 2\epsilon
     + \|\bu^*(\cdot+\br)-\bu^*(\bdot+\br')\|_2. $$
Since $\bu^*$ is smooth, we get
$$ \limsup_{\br'\rightarrow\br}
    \|\bu(\cdot+\br)-\bu(\bdot+\br')\|_2 \leq 2\epsilon. $$
Since $\epsilon>0$ is arbitrary, it follows that
$\lim_{\br'\rightarrow\br}\|\delta\bu(\br')\|_2=\|\delta\bu(\br)\|_2,$
completing the argument. To finish the proof, we observe by the triangle
and H\"{o}lder inequalities applied to (\ref{EMF-delta}) that
$$ \|\bepsilon_\ell\|_1 \leq
   \int d\br \,G_\ell(\br) \,\|\delta\bu(\br)\|_2\,\|\delta\bb(\br)\|_2
   + \int d\br \,G_\ell(\br)\, \|\delta\bu(\br)\|_2 \,
     \int d\br \,G_\ell(\br) \,\|\delta\bb(\br)\|_2.  $$
Since $\|\delta\bu(\br)\|_2$ is continuous and vanishes at $\br=\bzed,$
it follows that $\lim_{\ell\rightarrow 0}\|\bepsilon_\ell\|_1=0.$ QED

\vspace{4pt}

This result is in sharp contrast to that for the analogous ``vortex force''
in the hydrodynamic case, $\bF_\ell\equiv \ol{(\bu\btimes\bomega)}_\ell
-\ol{\bu}_\ell \btimes\ol{\bomega}_\ell,$ which does not need to
vanish in the limit as $\ell\rightarrow 0,$ even if the velocity
$\bu$ is continuous. It was proved in \cite{Eyink06a} that $\bF_\ell$
only needs to vanish if $\bu$ is H\"{o}lder continuous of order
greater than $1/2$. The Proposition 1 tells us that the limit of the EMF
along a subsequence for $\ell\rightarrow 0$ vanishes except on a set
of Lebesgue measure zero, when the total energy is finite.
However, this result does not imply conservation of flux for
every curve $C,$ since such sets have (three-dimensional) Lebesgue
measure zero and the line-integral of $\bepsilon_\ell$ on certain
choices of the loop $C$ might not vanish in the limit.

To get a result on flux conservation, let us prove a spatially
local version of the previous proposition, under stronger assumptions:
\begin{Prop}
If either $\bu$ or $\bb$ is continuous at point $\bx$ and if the
other is bounded, then $\bepsilon_\ell(\bx) \rightarrow 0$ as
$\ell\rightarrow 0.$
\end{Prop}

{\it Proof of Proposition 2:} Without loss of generality, let us
assume that $\bu$ is continuous at $\bx$ and that $|\bb(\bx')|
\leq B$ for $\bx'\in \Lambda$. Then, $|\delta\bb(\br;\bx)|\leq 2B$
and by the triangle inequality
\begin{eqnarray*}|
\bepsilon_\ell(\bx)| & \leq &
   \int d\br \,G_\ell(\br) \,|\delta\bu(\br;\bx)|\,|\delta\bb(\br;\bx)|
   + \int d\br \,G_\ell(\br)\, |\delta\bu(\br;\bx)| \,
     \int d\br \,G_\ell(\br) \,|\delta\bb(\br;\bx)| \cr
    & \leq & 4B \int d\br \,G_\ell(\br) \,|\delta\bu(\br;\bx)|.
\end{eqnarray*}
It follows then that $\lim_{\ell\rightarrow 0}\bepsilon_\ell(\bx)=\bzed,$
by continuity of $\bu$ at $\bx.$ Note that, if the filter kernel $G$
is compactly supported in space, then we need only assume that $\bb$
is bounded in a small neighborhood of $\bx.$ QED

\vspace{4pt}

\noindent This result has the very important implication that,
in order to get a non-vanishing EMF in the limit $\ell\rightarrow 0$
at a point $\bx$, {\it both} the velocity {\it and} the magnetic
field must be irregular there, at least discontinuous or even unbounded.

We can now deduce the following simple consequence for
flux conservation:
\begin{Cor}
Let $C$ be a closed, oriented, and rectifiable curve, and let
$\bu,\bb$ be bounded functions in a neighborhood of $C,$
such that at least one of them is continuous at every point
of $C$ except for a set of one-dimensional Hausdorff measure
$H^1$ equal to zero. Then,
$$ \lim_{\ell\rightarrow 0}\oint_{C}\bepsilon_\ell\bdot d\bx =0. $$
\end{Cor}
An equivalent statement of this result is that, for Alfv\'{e}n's
Theorem to break down, one of the following conditions must hold:
either (i) the curve $C$ must be non-rectifiable, or (ii) at least
one of $\bu$ or $\bb$ must be unbounded on $C,$ or else (iii) the
curve $C$ and the set of discontinuities ${\mathcal D}={\mathcal D}_\bu
\cap {\mathcal D}_\bb$ of both $\bu$ and $\bb$ must intersect in
a set of finite length. More technically, it is required that
\be    H^1(C\cap {\mathcal D})>0, \lb{hausdorff} \ee
where $H^1$ is the one-dimensional Hausdorff measure on subsets of
$\mathbb{R}^3$.

We shall return later to conditions (i) and (ii) in our discussion
of turbulent MHD flows. Here we focus on condition (iii), showing
by an example that it can indeed lead to violation of flux-conservation:
\begin{Ex}
Let
$$\bu(\bx) = \frac{1}{2}\Delta u_0 \,\sign(y)\, \hat{\bi}$$
$$ \bb(\bx) = \frac{1}{2}\Delta b_0 \,\sign(x\cos\varphi+y\sin\varphi)
         \, [\hat{\bj}\cos\varphi-\hat{\bi}\sin\varphi] $$
and let $G$ be any spherically symmetric (or even
cylindrically symmetric) filter kernel. Then, on the
$z$-axis
\be \bepsilon_\ell(0,0,z)=\frac{\Delta u_0 \Delta b_0}{2\pi}
             \sigma(\varphi)\cos(\varphi)\,\hat{\bk},
\lb{epsilon-z} \ee
independent of $\ell,$ while at all other points
$\lim_{\ell\rightarrow 0}\bepsilon_\ell(\bx)=0.$ Here
$\sigma$ is the $2\pi$-periodic function defined by
$$ \sigma(\varphi) =\left\{\begin{array}{ll}
                          \varphi & -\pi/2<\varphi<\pi/2 \cr
                         \pi-\varphi & \pi/2<\varphi<3\pi/2
                          \end{array} \right. $$

\end{Ex}

\begin{figure}
\epsfig{figure= 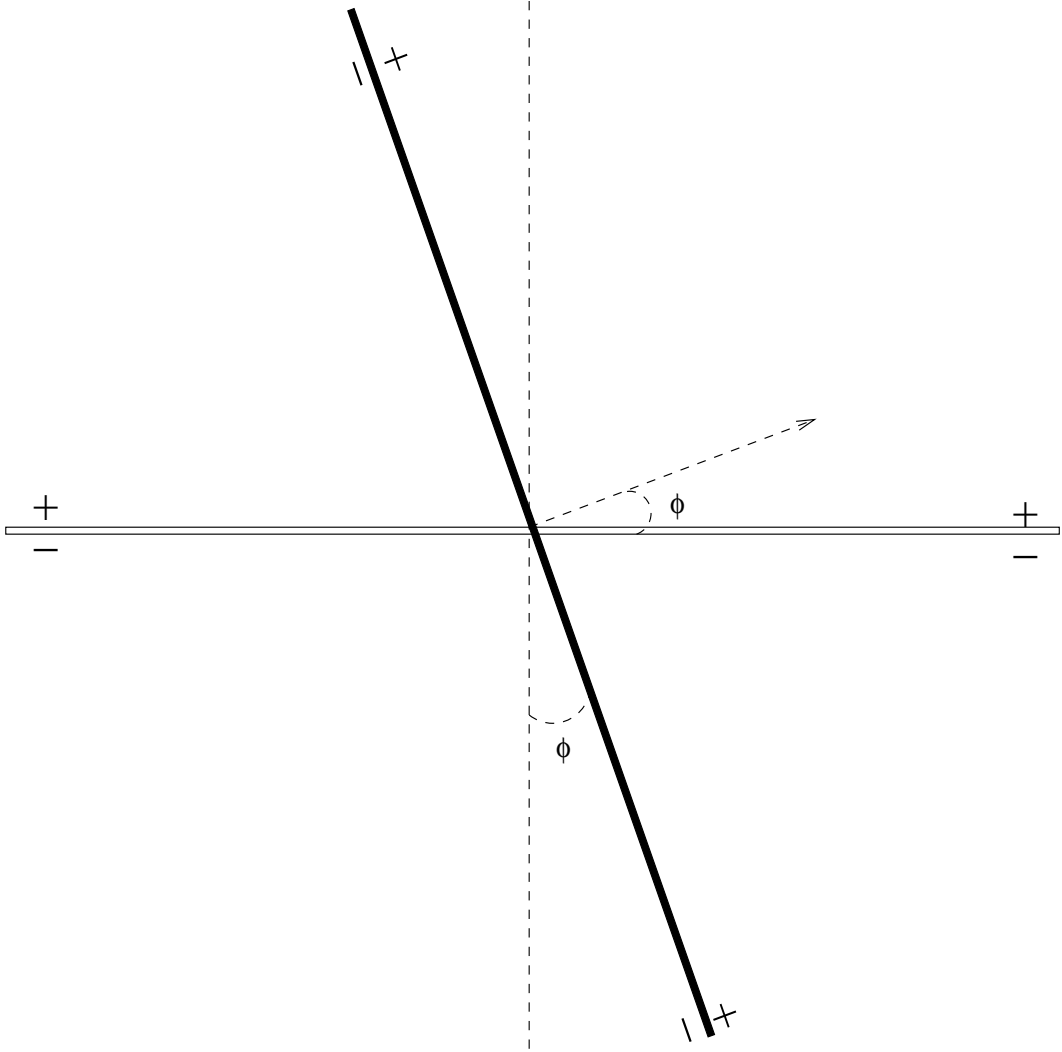,width=400pt,height=400pt}
\caption{An intersecting vortex sheet and current sheet viewed
along the axis of intersection (the $z$-axis). The white
strip represents the vortex sheet in the $xz$-plane. The
black strip represents the current sheet, in the plane
obtained by rotating the $yz$-plane by angle $\varphi$
around the $z$-axis. The $\pm$ labels on the sides of the
sheets indicate the values of the sign functions in the
definition of the velocity and magnetic field for this
example.}
\lb{Figure1}
\end{figure}

This example consists of a vortex sheet and a current sheet
intersecting in a line, with an angle of $\pi/2-\varphi$
between them. The vortex sheet has strength $\Delta u_0$
and lies in the $xz$-plane, while the current sheet has
strength $\Delta b_0$ and lies in a plane obtained by
rotating the $yz$-plane around the $z$-axis by angle $\varphi$.
(See Figure 1.) To establish (\ref{epsilon-z}), note first that
for any cylindrically symmetric filter $G$ the filtered
functions $\ol{\bu}_\ell(\bx),\ol{\bb}_\ell(\bx)$ vanish on the
$z$-axis. (In fact, they vanish on the entire sheets of
discontinuity of $\bu$ and $\bb,$ respectively.) Thus,
$\bepsilon_\ell=\ol{(\bu\btimes\bb)}_\ell$ on the $z$-axis.
Furthermore, in cylindrical coordinates $(r,\theta,z),$
$$ \bu(\bx)\btimes \bb(\bx) =  \frac{1}{4}\Delta u_0\,
  \Delta b_0 \,\sign(\sin(\theta))\sign(\cos(\theta-\varphi))
   \cos(\varphi) \hat{\bk}. $$
Thus, for any cylindrically-symmetric filter (not depending
upon $\theta$),
$$ \bepsilon_\ell(0,0,z)= \frac{1}{8\pi}\Delta u_0\,\Delta b_0
    \cos(\varphi) \hat{\bk}
 \int_0^{2\pi} d\theta \,\sign(\sin(\theta))\sign(\cos(\theta-\varphi)).
$$
The integral has the value $4\sigma(\varphi),$ giving the result
(\ref{epsilon-z}). Note that $\lim_{\ell\rightarrow 0}\bepsilon_\ell(\bx)
=0$ off the $z$-axis by our Proposition 2. In fact, this is
easy to see directly using the constancy of $\bu$ and $\bb$ off
the sheets, so that $\delta\bu=\delta\bb=\bzed,$  and formula
(\ref{EMF-delta}).

In the above example, consider a smooth loop $C$ that
has one segment consisting of an interval along the $z$-axis
of length $L_z$. Then for any such loop $C,$
$$ \lim_{\ell\rightarrow 0}\oint_{C}\bepsilon_\ell\bdot d\bx
   = \frac{\Delta u_0 \Delta b_0}{2\pi} L_z
             \sigma(\varphi)\cos(\varphi),  $$
if the orientation of the curve is upward along the segment
on the $z$-axis. The function $\sigma(\varphi)$ resembles
the trigometric sine function, but is piecewise linear,
and its product with $\cos(\varphi)$ is nonzero except for
$\varphi$ an integer multiple of $\pi/2.$ The maximum value
in the first quadrant, $0<\varphi<\pi/2,$ occurs for the
solution of $\cot\varphi_*=\varphi_*,$ or $\varphi_*\doteq
0.8603,$ an angle a bit larger than $\pi/4.$ In any case,
for any angle $\varphi$ not an integer mutiple of $\pi/2$ and
for an infinite set of loops, flux through the loop $C$ is not
conserved, instantaneously, in the limit as $\ell\rightarrow 0.$

\section{Physical Breakdown of Alfv\'{e}n's Theorem?}

In our discussion thus far we have used only the homogeneous
Maxwell equations (\ref{homo-Maxwell}) and the ideal Ohm's law
(\ref{ideal-Ohm}) and, in particular, their consequence, equation
(\ref{ideal-Faraday}), which we repeat here, for convenience, in
a somewhat different form:
\be \partial\bb/\partial t +(\bu\bdot\grad)\bb = (\bb\bdot\grad)\bu.
     \lb{ideal-Faraday2} \ee
To fully describe an ideal plasma, there must be adjoined also the
momentum equation:
\be \partial\bu/\partial t +(\bu\bdot\grad)\bu = (\bb\bdot\grad)\bb
                                       -\grad \tilde{p} \lb{momentum} \ee
where $\tilde{p}=p+b^2/2$ combines the hydrodynamic and magnetic
pressure. We now examine the possibilities for the breakdown of
Alfv\'{e}n's Theorem in the context of (\ref{ideal-Faraday2}) and
(\ref{momentum}), the equations of ideal magnetohydrodynamics.

Our Proposition 2 shows that discontinuous solutions are necessary for
the breakdown of flux conservation (or, even worse, unbounded
solutions). Of course, it is well-known that the ideal MHD equations
possess solutions that are piecewise smooth with jump discontinuities
on a smooth surface $\cD$. A compressible plasma possesses a richer set
of such solutions (including shocks), but here we restrict ourselves to
incompressible fluids. In that case the jump conditions at the surface
$\cD$ of discontinuity are \cite{Ilinetal03,LandauLifshitzECM}:
\be \Delta u_n=\Delta b_n=0 \lb{normal} \ee
\be b_n\Delta\bu_t=(u_n-v_n)\Delta\bb_t \lb{b-tangent} \ee
\be (u_n-v_n)\Delta \bu_t= b_n\Delta\bb_t,\,\,\,\, \Delta \tilde{p}=0
       \lb{u-tangent} \ee
Here $\Delta f$ for any quantity $f$ denotes its jump across $\cD.$ $u_n,b_n$
are the components of $\bu,\bb$ locally normal to the surface and they
must have no discontinuity, by the divergence-free condition. The
discontinuities
in the tangential components $\bu_t,\bb_t$ are related by (\ref{b-tangent}),
(\ref{u-tangent}), which follow from (\ref{ideal-Faraday2}),(\ref{momentum}).
Here $v_n$ denotes the velocity of the surface $\cD$ normal to itself.
These equations imply that $|u_n-v_n|=|b_n|$ and thus allow two classes
of solutions \cite{Ilinetal03,LandauLifshitzECM}. The first class has non-zero
mass flow across the surface $\cD$ of discontinuity:
\be |u_n-v_n|=|b_n|\neq 0 \Longrightarrow \Delta\bu_t = \pm\Delta\bb_t.
       \lb{mass-flow} \ee
The second type has no mass flow across $\cD$:
\be |u_n-v_n|=|b_n|=0 \Longrightarrow \Delta\bu_t,\,\Delta\bb_t\,\,\,\,
                                          {\rm arbitrary} \lb{tangent-D} \ee
The second type is called generally a  ``tangential discontinuity'', or, if
both
$\Delta\bu_t$ and $\Delta\bb_t$ are non-zero, a {\it current-vortex sheet}.

It is interesting to observe that a single such structure, in isolation, can
lead
to no breakdown of Alfv\'{e}n's Theorem. If the filter kernel $G$ is
spherically
symmetric, then it is easy to see that  at a point $\bx\in \cD,$
$$ \lim_{\ell \rightarrow 0}\ol{\bu}_\ell(\bx)=
\frac{1}{2}[\bu_+(\bx)+\bu_-(\bx)], $$
where $\bu_+(\bx),\bu_-(\bx)$ are the values of velocity $\bu$ approaching $\bx
$
from either side of $\cD$.  Using the similar results for
$\ol{\bb}_\ell,\,\ol{(\bu\btimes\bb)}_\ell$
gives immediately that
\be  \lim_{\ell\rightarrow 0}\bepsilon_\ell(\bx)
               =\frac{1}{4}\Delta\bu_t(\bx)\btimes\Delta\bb_t(\bx)
\lb{epsilon-tangent} \ee
for $\bx\in \cD.$   Of course, the limit is zero for $\bx\notin \cD.$ For the
type of
discontinuity with mass-flow, condition (\ref{mass-flow}) implies that the
limit
is zero also on $\cD.$ The second type of discontinuity, however, a general
current-vortex
sheet, can have a nonzero limit for $\bepsilon_\ell$ over the entire
two-dimensional
surface $\cD.$ Nevertheless, this EMF cannot lead to a violation of
Alfv\'{e}n's Theorem
in the limit, because it is everywhere normal to the surface and no
non-vanishing
line-integral is possible. Thus, our Example 1 of the preceding section, with a
pair
of intersecting sheets, seems  the simplest possibility for breakdown of flux
conservation.

Both current sheets and vortex sheets are commonly observed in numerical
simulations of near-ideal MHD equations, for both two-dimensions (2D) and
three-dimensions (3D).  We know of no evidence from these simulations
for any worse singularity with unbounded $|\bu|$ or $|\bb|.$ This seems to
indicate
that the condition (ii) of our Corollary 1 is only an academic possibility, not
physically realized. We shall review here briefly some of the available
numerical
results, with no attempt at completeness.

Current sheets and vortex sheets have been observed to develop in a variety
of simulations of freely-decaying 2D-MHD, in approximately ideal conditions.
Initial conditions that have been employed include the Orszag-Tang vortex
or slight modifications \cite{Frischetal83,BiskampWelter89,GrauerMarliani98}
and random initial conditions \cite{BiskampWelter89}. The current sheets and
vortex sheets appear usually in very close proximity. A rather successful
analytical theory was developed in \cite{Sulemetal85} for the formation of
these structures near $X$-type magnetic null points. Note, however, that
there can be no inviscid breakdown of Alfven's Theorem in 2D, from current
sheets and vortex sheets, or from any other type of singularity. Indeed, by its
definition, the subscale EMF vector $\bepsilon_\ell$ points always normal to
the
2D plane and therefore no non-vanishing line integral is possible. This is in
contrast to the hydrodynamic case, where a breakdown of the Kelvin Theorem
is possible in 2D \cite{Eyink06a,Eyink06b,Chenetal06}.

Current sheets and vortex sheets are also observed to develop in simulations
of 3D-MHD, at moderate magnetic Reynolds numbers. These have been observed
in freely-evolving flows initialized by a 3D extension of the Orszag-Tang
vortex
\cite{Politanoetal95,GrauerMarliani00}, by linked flux rings
\cite{GrauerMarliani00},
or by random initial conditions \cite{Politanoetal95}, and also in forced 3D
MHD
turbulence \cite{MaronGoldreich01}. The latter work \cite{MaronGoldreich01} did
not study $\bu,\,\bb$ fields but instead Elsasser variables
$\bz^\pm=\bu\pm\bb,$
further decomposed into contributions from shear-Alfv\'{e}n and
pseudo-Alfv\'{e}n
modes. It was found that all four of these fields form strong sheet-like
singularities
in close proximity. Of course, the subscale EMF may be written also in terms
of Elsasser fields, as
\be \bepsilon_\ell =\frac{1}{2}\left[\ol{(\bz^-\btimes\bz^+)}_\ell
-\ol{\bz}^{\,-}_\ell\btimes\ol{\bz}^{\,+}_\ell\right], \ee
and intersecting sheets of discontinuity of $\bz^+$ and $\bz^-$ can  generate
a non-vanishing EMF.  The simulations in \cite{Politanoetal95} found closely
associated sheets for $\bu,\,\bb$ and they concluded that  ``vorticity and
current
display similar features and are usually intense in adjacent regions.'' This
paper
also studied the dynamics of these structures, their formation and persistence.
The latter is an important issue, since intersection of vortex and current
sheets
are required, not only at an instant  but also over some interval of time. By
equation (\ref{filter-Alfven}), the large-scale magnetic fluxes at two
subsequent times $t'>t$ are related by
\be \ol{\Phi}_\ell(S,t') -  \ol{\Phi}_\ell(S,t) =
       \int_t^{t'} d\tau \oint_{\ol{C}_\ell(\tau)} \bepsilon_\ell(\tau)\bdot
d\bx.  \lb{} \ee
(if all non-ideal effects may be ignored at scale $\ell$.) Thus, the magnetic
flux will
be conserved if the line-integral of the EMF is non-zero only for a set of
times of
measure zero. Our Example 1 in the previous section does not address this
issue,
because the configuration of intersecting vortex and current sheets employed
there
is not a stationary solution of the ideal MHD equations.

Nevertheless, we conjecture that magnetic flux conservation may indeed be
broken in ideal MHD by nonlinear effects.  As a physical phenomenon, it should
be analogous to the decay of magnetic flux trapped in a narrow superconducting
ring.
According to both theory
\cite{LangerAmbegaokar67,McCumberHalperin70,Zaikinetal97}
and experiment \cite{Newboweretal72,Lauetal01} this decay is due
to nucleation of quantized magnetic flux lines (by thermal, quantum, or other
fluctuations), which locally destroy the superconducting state. The quantized
flux
lines migrate out of the ring, allowing the relative phase across the point of
escape
to slip by $2\pi$ and generating a voltage pulse around the ring. Here it is
important that the quantized flux lines need not move with the local superfluid
velocity, due primarily to drag forces generated by their interaction with the
background excitations (quasi-particles and holes) \cite{AoZhu99}. The
physics in ideal MHD is similar, with the necessary singularities provided
by the (intersections of) current sheets and vortex sheets. In the presence
of such singularities, the large-scale magnetic field lines do not move with
the plasma velocity at the same scale but  gain a ``slip velocity'' due to
their
interaction with the subscale modes: see eq. (\ref{drift-velocity}). The
diffusion
of the lines of force of the large-scale magnetic field out of the advected
loop
implies a voltage pulse around the loop, which can lead to violation of flux
conservation. There is not only a physical similarity of this process with
quantum-phase slip in superconductors but also, as we discussed in
\cite{Eyink06b}, a fairly close formal analogy as well.

To complete this section we would like to make a few comments on the
relation of our results to various theories of magnetic line reconnection
in MHD.  Our simple Example 1 provides conditions similar to those
required in several such theories. In quasi-2D reconnection, there
is an $X$-type magnetic neutral line along which the parallel component
of the electric field is non-vanishing \cite{Axford84}. Such neutral lines are
not
structurally stable in 3D, so theories of 3D reconnection are often based
instead
upon neutral points (magnetic nulls) which are stable
\cite{Greene88,LauFinn90}.
These theories require a non-vanishing line-integral of the electric field
along magnetic field lines that connect pairs of nulls (null-null lines).
Finally, theories of $\bB\neq\bzed$ reconnection
\cite{Schindleretal88,HesseSchindler88,Schindler95} are based upon
magnetic field-lines of maximum loop voltage (eq.(\ref{loop-voltage})). In our
Example 1,  the current sheet is a neutral sheet for the large-scale magnetic
field $\ol{\bb}_\ell$, for any  $\ell>0.$ In particular, a neutral line
exists along which there is a non-vanishing integral of the electric field.
Adding a smooth external magnetic field with only a $z$-component, and
non-vanishing
on the $z$-axis, makes this null line a magnetic field line with  maximal
loop-voltage in the limit $\ell\rightarrow 0.$ By adding an appropriate
smooth external magnetic field, this line can also be converted to a null-null
field line. However, all of these constructions are contrived and clearly
inadequate
as a general model of fast reconnection (i.e. with rates independent of
resistivity).
In the next section we shall indicate what we believe are some missing
ingredients of such a theory.

The most important implication for theories of fast reconnection follows
from Proposition 2 and its Corollary 1. These show that vortex sheets
are equally important as current sheets to obtain non-vanishing reconnection
in the ideal limit. Any successful theory must involve essentially the
coincident
singularities of the velocity field $\bu$ and magnetic field $\bb.$

\section{Turbulent Cascade of Magnetic Flux}

The results in the present paper are not specific to turbulent plasma flows
but this is one of their most interesting areas of application. The necessary
ingredients to violate Alfv\'{e}n's Theorem---singular current and vortex
sheets
---seem to exist in MHD turbulence in the limit of high fluid and magnetic
Reynolds numbers. Therefore,  we expect that magnetic-flux conservation breaks
down under turbulent conditions. Analogous to the case of fluid velocity
circulations discussed in \cite{Eyink06a,Eyink06b,Chenetal06}, we may term
this a ``cascade of magnetic-flux''. However, the term ``cascade'' is not as
well warranted here, since the scale-locality of the process is in serious
doubt. Following the discussion in \cite{Eyink06b,Eyink05}, the turbulent
EMF $\bepsilon_\ell(\bx)$ is infrared (IR) local-in-scale if the H\"{o}lder
exponents of $\bu,\bb$ at the point $\bx$ satisfy $h_u<1$  and $h_b<1.$
Similarly, the turbulent EMF is ultraviolet (UV) local-in-scale at the point
$\bx$
if the H\"{o}lder exponents there satisfy $h_u>0$ and $h_b>0.$ This means that
the EMF is UV-local away from discontinuities. However it is precisely due to
these points that the flux-conservation is violated! It is possible that
UV-locality
still holds at such points, but it requires extensive cancellations between
the contributions from length-scales $\ll \ell.$ This is unlikely if the
current
sheets and vortex sheets are highly coherent structures at all length-scales.
Thus, it is more likely that UV locality is only marginal there. This
complicates
the task of developing adequate theoretical models for the turbulent EMF
$\bepsilon_\ell(\bx).$ For example, the ``multi-scale gradient'' expansion
that was developed in \cite{Eyink06c} and applied to the circulation cascade
in \cite{Eyink06b} is based upon UV-locality. Its application to the cascade
of magnetic-flux in MHD turbulence may thus be only qualitatively successful
for $\ell\rightarrow 0.$

Another complication in turbulent MHD flows is that material curves $C(t)$
advected by a velocity field which is not differentiable but only H\"{o}lder
continuous are expected to become fractal, with a Hausdorff dimension $>1$
\cite{Mandelbrot76,SreenivasanMeneveau86}. This is likely to occur in MHD
turbulence in the limit of infinite magnetic and fluid Reynolds numbers.
Fractality of material curves provides another potential mechanism for
breakdown of Alfv\'{e}n's Theorem, since fractal curves are non-rectifiable
 (condition (i) in Corollary 1.) In fact, for fractal curves and surfaces it is
not
 even clear how to {\it define} integrals such as the magnetic-flux
(\ref{flux})
 or the loop-voltage (\ref{Alfven-flux}). One possibility is to write, for
example,
 \be \oint_{C(t)}\, \bR(\bx,t)\bdot d\bx = \oint_C \bR_L(\ba,t) \bdot
d\bx(\ba,t)
  \lb{Stieltjes} \ee
where $\bx(\ba,t)$ is the Lagrangian flow map, defined by
$$ (d/dt)\bx(\ba,t)=\bu(\bx(\ba,t),t),\,\,\,\,\bx(\ba,t_0)=\ba, $$
$\bR_L(\ba,t)=\bR(\bx(\ba,t),t)$ is the non-ideality in a Lagrangian frame,
and the integral on the right side of (\ref{Stieltjes}) is defined over the
initial loop
$C$ at time $t_0.$ It was shown by Young \cite{Young36,Young37}
that this integral exists in the Stieltjes-sense if the minimal
H\"{o}lder exponents $h_R$ of $\bR_L(\ba,t)$ and $h_x$ of $\bx(\ba,t)$
satisfy $h_R+h_x>1.$ We expect that condition (i) can be removed
in Corollary 1 by an application of such ideas. An interesting laboratory
in which to study this question is the Kazantsev-Kraichnan dynamo
model \cite{Kazantsev68,Kraichnan68}, for the case of non-smooth
advecting velocity field \cite{Vincenzi02,BoldyrevCattaneo04,Celanietal06}.
It is expected that advected curves and surfaces in this model will become
fractal, just as in real turbulence \cite{Eyink06a}. However, the advecting
random velocity field is Gaussian and monofractal, so that there are no
vortex sheets (and perhaps no current sheets). Thus, the effect of fractality
of material objects can be studied in isolation. We conjecture that the
magnetic flux is strictly conserved for {\it all} surfaces in the
Kazantsev-Kraichnan
model, when the advecting velocity is non-smooth but H\"{o}lder continuous.

Although we expect no direct effect of fractality of the surface $S(t)$ (or of
its boundary $C(t)$) on conservation of magnetic flux, there can be an indirect
effect. If the Hausdorff dimension of $C(t)$ is $>1$, then it increases the
probability of a nontrivial intersection of the loop with the discontinuity set
$\cD$ of $\bu$ and $\bb$ (cf. eq.(\ref{hausdorff})).  If the Hausdorff
dimensions
of the current-vortex sheets are $>2$, then the dimension of their typical
intersection $\cD$ will be $>1$ and this will also enhance the probability of
condition (\ref{hausdorff}) being satisfied. Most phenomenological models
of intermittency in MHD turbulence have assumed that the Hausdorff
dimension of the sheets is exactly equal to 2
\cite{Graueretal94,PolitanoPouquet95,HorburyBalogh97,MullerBiskamp00}.
However, it is plausible to expect that turbulent advection on all scales will
lead
to wrinkling of the sheets, increasing their dimensionality to values $>2.$

A similar effect will appear due to the  {\it spontaneous stochasticity} of
magnetic field-lines in the limit of infinite magnetic Reynolds numbers. Field
lines $\bxi(\sigma,t)$ are defined in principle at each time $t$ by solving
the ODE (in the parameter $\sigma$ related to arclength $s$ by $ds
=|\bb|d\sigma$)
\be (d/d\sigma)\bxi(\sigma,t)=\bb(\bxi(\sigma,t),t), \,\,\,\,
                           \bxi(0,t)=\bxi_0 \lb{field-line} \ee
for the given magnetic field $\bb(\bx,t).$ However, in the limit of infinite
magnetic Reynolds number, the magnetic field is non-smooth and, in fact,
probably nowhere-differentiable. In that case, the solutions of
(\ref{field-line})
are not only fractal but also presumably random. This line-stochasticity can
arise
mathematically from the non-uniqueness of the solutions of the initial-value
problem
(\ref{field-line}) when $\bb$ is non-smooth. Physically, it corresponds to a
turbulent
diffusion in the arclength parameter $s,$ analogous to Richardson diffusion of
material
particles in hydrodynamic turbulence \cite{Richardson26}. This  phenomenon
of ``spontaneous stochasticity'' was first noted for Lagrangian trajectories in
the Kraichnan model of random advection  \cite{Bernardetal98,Chavesetal03}.
It has since been rigorously proved in the Kraichnan model that the solutions
for the Lagrangian trajectories correspond to a random process, with a
{\it fixed} initial condition $\bx_0$ for the fluid particle and a {\it fixed}
advecting
velocity $\bu$ \cite{LeJanRaimond02,LeJanRaimond04}. These considerations
carry over plausibly also to the equation (\ref{field-line}) for the magnetic
field-lines.
Such stochastic effects will increase the likelihood of magnetic field-lines
intersecting
the singular set $\cD.$ Note that this type of random field-line wandering is a
crucial
part of current theories of fast turbulent reconnection
\cite{LazarianVishniac99,LazarianVishniac00}.

\section{Conclusions}

The results presented in this paper support theories of fast turbulent
reconnection \cite{LazarianVishniac99,LazarianVishniac00,Lazarian05},
in a general way, but also place rigorous constraints upon them. A basic
assumption of those theories is that Alfv\'{e}n's Theorem may be violated
in the limit of vanishing resistivity. We have shown that this is possible
by an analysis  of the MHD equations for an ideal plasma. However, in
contrast to Kelvin's circulation theorem, which is rather easily violated
\cite{Eyink06a,Eyink06b,Chenetal06}, Alfv\'{e}n's theorem on magnetic-flux
conservation is much more robust. We have proved that violations of it are
only possible, essentially, if singular vortex sheets and current sheets
have intersections with high enough dimension and persist long enough in
time.  These results should help to guide further theoretical, numerical,
and experimental work.

We have shown, in particular, that it is crucial to understand the physical
properties of the subscale EMF, defined by eq.(\ref{filter-EMF}). At large
enough length-scales $\ell$, this quantity is the critical driver of magnetic
line reconnection and any non-ideality at small scales is irrelevant.
To explain the fast rates of reconnection observed in astrophysical
situations, where many decades of inertial range are often observed,
a quantitative theory for the turbulent EMF must be developed.

The predicted breakdown of magnetic-flux conservation in ideal MHD,
as a physical phenomenon, is closely analogous to the decay by quantum
phase-slip of magnetic flux confined in a superconducting ring. It should
be observable both in numerical simulations and in laboratory experiments
at moderately high Reynolds numbers.

\vspace{.1in}
\noindent {\small
{\bf Acknowledgements.} We wish to thank E. T. Vishniac for many illuminating
discussions on the subject of this paper. This work was supported by NSF
grant \# ASE-0428325 at the Johns Hopkins University and by the Center
for Nonlinear Studies at Los Alamos National Laboratory.}

\end{document}